\documentclass[aps,amsfonts,amsmath,prd,preprint,nofootinbib]{revtex4}

\usepackage[utf8]{inputenc}
\usepackage{amsfonts}
\usepackage{physics}
\usepackage{mathtools}
\usepackage{mathrsfs}
\usepackage{amssymb}
\usepackage{appendix}
\usepackage{graphicx}
\usepackage{amsmath}
\usepackage{color}
\usepackage{feynmf}
\usepackage{simpler-wick} 
\usepackage[legalpaper, margin=1in]{geometry}

\begin{document}

\title{A Summation of Series Involving Bessel Functions and Order Derivatives of Bessel Functions}

\author{Yilin Chen}
\email{Yilin.Chen@tufts.edu}
\address{Institute of Cosmology, Department of Physics and Astronomy, \\
Tufts University, Medford, Massachusetts 02155, USA}

\begin{abstract}
    In this note, we derive the closed-form expression for the summation of series $\sum_{n=0}^{\infty}nJ_n(x)\partial J_n/\partial n$, which is found in the calculation of entanglement entropy in 2-d bosonic free field, in terms of $Y_0$, $J_0$ and an integral involving these two Bessel functions. Further, we point out the integral can be expressed as a Meijer G function.
\end{abstract}
\maketitle

\section{Introduction}
In this note, we find that the sum of a series involving Bessel functions and the order derivatives of Bessel functions:
\begin{align}
    P(x)\equiv\sum_{n=0}^{\infty}nJ_n(x)\frac{\partial J_{n}(x)}{\partial n}
\end{align}
can be expressed in a closed-form.

The motivation of finding this sum rule arises from the study of entanglement entropy in two dimensional bosonic free field. In \cite{Calabrese_2004}, the authors provide a way to compute such entanglement entropy. The key point is the Green's function in the $n$-sheeted geometry $G_n(\mathbf{r},\mathbf{r})$. It obeys the Helmholtz equation in polar coordinates and can be expressed as
\begin{align}\label{green}
    G_n(r,\theta;m^2)=\frac{1}{2n\pi}\sum_{k=0}^{\infty}\varepsilon_k\int_0^{\infty}\lambda d\lambda \frac{J^2_{k/n}(\lambda r)}{\lambda^2+m^2},
\end{align}
where $\varepsilon_0=1$ or $\varepsilon_n=2$ otherwise. With the Green's function in the $n$-sheeted geometry, the R\'enyi entropy can be expressed as
\begin{align}
    \Tr\rho^n=\exp{-\frac{1}{2}\int_{a^2}^{m^2}dm^{'2}\int d^2r \left(G_n(r,\theta;m^{'2})-nG_1(r,\theta;m^{'2})\right)}.
\end{align}
Then the von Neumann entropy
\begin{align}
    S=-\Tr\rho\log\rho=\left.-\frac{\partial}{\partial n}\Tr\rho^n\right|_{n=1}=\frac{1}{2}\int_{a^2}^{m^2}dm^{'2}\int d^2r \left(\left.\frac{\partial}{\partial n}G_n(r,\theta;m^{'2})\right|_{n=1}-G_1(r,\theta;m^{'2})\right).
\end{align}
It is not hard to see the summation over the product of Bessel functions in \eqref{green} playing an important role. In the last step calculating the von Neumann entropy, the derivative with respect to $n$ requires the \textit{order derivatives} and the summation now becomes
\begin{align}
    \left.\frac{\partial}{\partial n}\sum_{k=0}^{\infty}\varepsilon_k J^2_{k/n}(\lambda r)\right|_{n=1}=\left.2\sum_{k=0}^{\infty}\varepsilon_k \frac{-k}{n^2}J_{k/n}(\lambda r)\frac{\partial J_{k/n}(\lambda r)}{\partial (k/n)}\right|_{n=1}=-4P(\lambda r).
\end{align}
\section{Derivation of $P(x)$}
Let us start with this formula found in \cite{Order}
\begin{align}
    \int^{\infty}_x\frac{J^2_{\nu}(t)}{t}dt=\frac{1}{2\nu}-\frac{1}{2\nu}\sum_{n=0}^{\infty}\varepsilon_n J^2_{\nu+n}(x),
\end{align}
where $\varepsilon_0=1$ and $\varepsilon_n=1$ otherwise.
By multiplying $\nu$ on both sides and taking the derivative with respect to $\nu$, we have
\begin{align}\label{im}
    \int^{\infty}_x\frac{J^2_{\nu}(t)+2\nu J_{\nu}(t)\hat{J}_{\nu}(t)}{t}dt=-\sum_{n=0}^{\infty}\varepsilon_n J_{\nu+n}(x)\hat{J}_{\nu+n}(x)=-2\sum_{n=1}^{\infty}J_{\nu+n}(x)\hat{J}_{\nu+n}(x)-J_{\nu}(x)\hat{J}_{\nu}(x),
\end{align}
where $\hat{J}_{\nu}=\partial J_{\nu}/\partial\nu$.
Then we set $\nu\in\mathbb{N}$ and sum over $\nu$ form $0$ to $\infty$. \eqref{im} becomes
\begin{align}\label{imp}
    \int^{\infty}_x\frac{\sum_{\nu=0}^{\infty}J^2_{\nu}(t)+2P(t)}{t}dt+\sum^{\infty}_{\nu=0}J_{\nu}(x)\hat{J}_{\nu}(x)=-2\sum^{\infty}_{\nu=0}\sum^{\infty}_{n=1}J_{\nu+n}(x)\hat{J}_{\nu+n}(x).
\end{align}
Notice that the RHS gives $-2P(x)$. From (3.2) in \cite{Order}, the second term in the LHS of \eqref{imp} gives
\begin{align}
    \sum^{\infty}_{\nu=0}J_{\nu}(x)\hat{J}_{\nu}(x)=-\frac{1}{2}\int^{\infty}_x\frac{J^2_0(t)}{t}dt+\frac{1}{2}J_0(x)\hat{J}_0(x),
\end{align}
and also
\begin{align}
    \sum_{\nu=0}^{\infty}J^2_{\nu}(t)=\frac{1}{2}\left(\sum_{\nu=-\infty}^{\infty}J^2_{\nu}(t)+J^2_0(t)\right)=\frac{J^{2}_0(t)+1}{2}.
\end{align}
Thus, \eqref{imp} reduces to
\begin{align}
    \int^{\infty}_x\left(\frac{J^{2}_0(t)+1}{2}+2P(t)\right)\frac{dt}{t}-\frac{1}{2}\int^{\infty}_x\frac{J^2_0(t)}{t}dt+\frac{1}{2}J_0(x)\hat{J}_0(x)=-2P(x).
\end{align}
After taking the derivative with respect to $x$ and substituting $\hat{J}_0(x)=\frac{\pi}{2}Y_0(x)$ (8.486(1) in \cite{GR}), we have the differential equation
\begin{align}
    P'(x)-\frac{P(x)}{x}=\frac{1}{4x}-\frac{\pi}{8}(J_0(x)Y_0(x))'.
\end{align}
This equation can be easily solved by the method of variation of parameters. The solution
\begin{align}
    P(x)&=-x\left(\int^{\infty}_x\left(\frac{1}{4t^2}-\frac{\pi}{8}(J_0(t)Y_0(t))'\right)dt+C\right)\nonumber\\
    &=-x\left(\frac{1}{4x}-\left.\frac{\pi}{8t}J_0(t)Y_0(t)\right|_x^{\infty}-\frac{\pi}{8}\int^{\infty}_x\frac{J_0(t)Y_0(t)}{t^2}dt+C\right)\nonumber\\
    &=-\frac{1}{8}\left(2+\pi J_0(x)Y_0(x)-\pi x\int^{\infty}_x\frac{J_0(t)Y_0(t)}{t^2}dt+Cx\right).
\end{align}
To determine the integral constant $C$, we can analyse the asymptotic behavior of $P(x)$ at large $x$. When $x\to\infty$, we have $P(x)\sim-\frac{1}{8}(2+Cx)$, which means when $x$ is sufficiently large, $P(x)$ is approximately a linear function and the slope is $C$. However, the asymptotic form of $\hat{J}_n$ can be written as \cite{Order}
\begin{align}
    \hat{J}_{n}(x)\sim \sqrt{\frac{\pi}{2x}}\sin\left(x-\frac{n\pi}{2}-\frac{\pi}{4}\right)\ \ (x\to\infty).
\end{align}
Combining with the asymptotic approximation of $J_n$ in \eqref{as}, it is easy to find that $P(x)\sim \frac{a}{x}+b\sim b$ when $x\to\infty$, where $a$ and $b$ are constants. More detailed discussion about the asymptotic behavior can be found in the next section.
So the integral constant $C=0$ and finally we obtain a closed-form
\begin{align}\label{origin}
    P(x)=-\frac{1}{8}\left(2+\pi J_0(x)Y_0(x)-\pi x\int^{\infty}_x\frac{J_0(t)Y_0(t)}{t^2}dt\right).
\end{align}
Also, we can evaluate the integral (see Appendix), which can be expressed by a Meijer G function
\begin{align}
    \int^{\infty}_x\frac{J_0(t)Y_0(t)}{t^2}dt=-\frac{1}{2\sqrt{\pi}}G^{20}_{13}\left(x^2\middle\vert
    \begin{array}{c}
         1 \\
         -\frac{1}{2},-\frac{1}{2},-\frac{1}{2}
    \end{array}
    \right).
\end{align}
Hence, we have an another expression
\begin{align}
    P(x)=-\frac{1}{8}\left(2+\pi J_0(x)Y_0(x)+ \frac{\sqrt{\pi}x}{2}G^{20}_{13}\left(x^2\middle\vert
    \begin{array}{c}
         1 \\
         -\frac{1}{2},-\frac{1}{2},-\frac{1}{2}
    \end{array}
    \right)\right).
\end{align}
\section{Asymptotic Approximation of $P(x)$ and Regularization of Green's Functions}
We now obtain the asymptotic approximation of $P(x)$ at large $x$. The asymptotic behavior can be helpful to analyze the behavior of the combination of Green's functions $\left.\partial G_n(r,\theta;m^{'2})/\partial n\right|_{n=1}-G_1(r,\theta;m^{'2})$, especially the divergence and the regulariztion of it.

Obviously, when $x\to\infty$, the last two terms in \eqref{origin}, involving $J_0$ and $Y_0$, go to zero fast (at the order of $O(x^{-1})$). So we have
\begin{align}
    \lim_{x\to\infty}P(x)=-\frac{1}{4}.
\end{align}
This is not a good news for us because $P(x)$ as a part of the integrand in $\left.\partial G_n(r,\theta;m^{'2})/\partial n\right|_{n=1}$, the non-zero limits highly increases the probability that we have a divergent integral. Indeed, the integral is actually divergent.

However, things are not so bad. We define
\begin{align}
    Q(x)\equiv &-4P(x)-\sum_{k=0}^{\infty}J_k^2(x)=-4P(x)-1\nonumber\\
    &=\frac{1}{2}\left(\pi J_0(x)Y_0(x)-\pi x\int^{\infty}_x\frac{J_0(t)Y_0(t)}{t^2}dt\right).
\end{align}
We can rewrite the combination of Green's functions in terms of $Q(x)$:
\begin{align}\label{combination}
    \left.\frac{\partial}{\partial n}G_n(r,\theta;m^{'2})\right|_{n=1}-G_1(r,\theta;m^{'2})=\frac{1}{2\pi}\int_0^{\infty}\frac{Q(\lambda r)}{\lambda^2+m^{'2}}\lambda d\lambda.
\end{align}
It is easy to check that $Q(x)$ vanishes when $x$ approaches to infinity. By introducing asymptotic approximations for $J_0(x)$ and $Y_0(x)$ at large $x$
\begin{align}\label{as}
    J_n(x)\sim\sqrt{\frac{2}{\pi x}}\cos\left(x-\frac{n\pi}{2}-\frac{\pi}{4}\right),\nonumber\\
    Y_n(x)\sim\sqrt{\frac{2}{\pi x}}\sin\left(x-\frac{n\pi}{2}-\frac{\pi}{4}\right),
\end{align}
we have $\pi J_0(x)Y_0(x)\sim -\cos(2x)/x$. Substituting this approximation into the integral and integrating by part, we derive the asymptotic approximation for the second term in $Q(x)$, which gives
\begin{align}
    \int_x^{\infty}\frac{J_0(t)Y_0(t)}{t^2}dt\sim-\int_x^{\infty}\frac{\cos(2t)}{t^3}dt=\frac{\sin(x)}{2x^3}-\frac{3}{2}\int_x^{\infty}\frac{\sin(2t)}{t^4}dt\nonumber\\
    \sim O(x^{-3}).
\end{align}
Comparing to the first term, we can neglect the second term when $x$ is sufficiently large. Thus, the asymptotic approximation of $Q(x)$ turns
\begin{align}
    Q(x)\sim-\frac{\cos(2x)}{x}.
\end{align}
Also, we can evaluate the value at $x=0$. By using L'Hopital rule, we find
\begin{align}
    \lim_{x\to0}&\frac{1}{1/x}\left(\frac{J_0(x)Y_0(x)}{x}-\int^{\infty}_x\frac{J_0(t)Y_0(t)}{t^2}dt\right)\nonumber\\
    &=\lim_{x\to0}\frac{-1}{1/x^2}\left(\frac{x(J_0(x)Y_0(x))'-J_0(x)Y_0(x)}{x^2}+\frac{J_0(x)Y_0(x)}{x^2}\right)\nonumber\\
    &=-\lim_{x\to0}x(J_0(x)Y_0(x))'.
\end{align}
Using identities (8.471 and 8.477 in \cite{GR})
\begin{align}
    &Z'_n(x)=\frac{1}{2}(Z_{n-1}(x)-Z_{n+1}(x)),\nonumber\\
    &J_n(x)Y_{n+1}(x)-J_{n+1}(x)Y_{n}(x)=-\frac{2}{\pi x},\nonumber
\end{align}
where $Z$ denotes $J$ or $Y$, we have
\begin{align}
    -\lim_{x\to0}x(J_0(x)Y_0(x))'&=\lim_{x\to0}x(J_1(x)Y_0(x)+J_0(x)Y_1(x))\nonumber\\
    &=\lim_{x\to0}x\left(2J_1(x)Y_0(x)-\frac{2}{\pi x}\right)\nonumber\\
    &=-\frac{2}{\pi}.
\end{align}
Thus, $Q(0)=-1$. Moreover, no singularity at the origin (of course, no singularity at other place as well) and the simple asymptotic form of $Q(x)$ tell us the combination in \eqref{combination} is convergent. Also, by using the identity (7.811 5) in \cite{GR1} and the relation (9.31 1) in \cite{GR} to deal with the Meijer G function, one can evaluate the combination exactly
\begin{align}\label{final}
    \left.\frac{\partial}{\partial n}G_n(r,\theta;m^{'2})\right|_{n=1}-G_1(r,\theta;m^{'2})=\frac{1}{16\pi} \left(\sqrt{\pi }m' r G_{13}^{30}\left(m^{'2} r^2\middle\vert
\begin{array}{c}
 1 \\
 -\frac{1}{2},-\frac{1}{2},-\frac{1}{2} \\
\end{array}
\right)-4 K^2_0\left(m'r\right)\right).
\end{align}
Furthermore, we can also obtain
\begin{align}
    \int d^2r\left(\left.\frac{\partial}{\partial n}G_n(r,\theta;m^{'2})\right|_{n=1}-G_1(r,\theta;m^{'2})\right)=-\frac{1}{6 m^{'2}},
\end{align}
which agrees the result in \cite{Calabrese_2004}. Because the radial integral is definitely a Mellin transform so this result can be found in Mellin transform table in \cite{tables}. 
\section{Conclusions and Remarks}
We study a summation of series involving Bessel functions and the order derivatives of Bessel functions, which arises in the study of entanglement entropy in two dimensional bosonic free field. We find a closed-form expression of such summation which can be expressed simply in terms of $J_0$, $Y_0$ and a Meijer G function. Further, we study the asymptotic behavior of this summation of series and finally we find a simple expression for the combination of Green's functions in \eqref{final}.

The expression in \eqref{final} and also $P(x)$, $Q(x)$ might be useful for calculating entanglement entropies in different fields. For instance, if the field is coupled with a potential such that the equation of motion is no longer a homogeneous Helmholtz equation and the Green's equation becomes
\begin{align}
    (-\nabla^2_{\mathbf{r}}+m^2-f(\mathbf{r}))G_n(\mathbf{r},\mathbf{r'})=\delta(\mathbf{r}-\mathbf{r'}).
\end{align}
Generally, the exact solution would not be found easily. However, the equation can be solved perturbatively by applying the resolvent formalism
\begin{align}\label{resolvent}
    G_n(\mathbf{r},\mathbf{r'})=\Tilde{G}_0(\mathbf{r},\mathbf{r'})+\Tilde{G}_1(\mathbf{r},\mathbf{r'})+...,
\end{align}
where
\begin{align}\label{expansion}
    &\Tilde{G}_0(\mathbf{r},\mathbf{r'})=\Tilde{G}(\mathbf{r},\mathbf{r'}),\nonumber\\
    &\Tilde{G}_1(\mathbf{r},\mathbf{r'})=(\Tilde{G}\circ f\circ \Tilde{G})(\mathbf{r},\mathbf{r'}),\nonumber\\
    &...\nonumber\\
    &\Tilde{G}_i(\mathbf{r},\mathbf{r'})=(\Tilde{G}\circ f\circ \Tilde{G}\circ...\circ f\circ \Tilde{G})(\mathbf{r},\mathbf{r'}),
\end{align}
where $\Tilde{G}$ is the Green's function with $f=0$ and $(G_n\circ f)(\mathbf{r})=\int  G(\mathbf{r},\mathbf{r'})f(\mathbf{r'})d^2r$. In this method, the exact Green's function can be also written as (see Appendix B)
\begin{align}\label{full}
    G_n(\mathbf{r},\mathbf{r'})=\sum_{l=0}^{\infty}\sum_{i+j=0}^l B_{i,j,l}[f](2\partial_{\omega})^i(2\partial_{\bar{\omega}})^j \Tilde{G}_n^{\left(1+\frac{i+j+l}{2}\right)}(\mathbf{r},\mathbf{r'}).
\end{align}
This method is well developed in \cite{dikii1955zeta,dikii1958trace} for Sturm-Liouville problems. More details can be found in \cite{gelfand,SEELEY}. Furthermore, we can apply the Green's function to calculate the entanglement entropy
\begin{align}
    &\left.\frac{\partial}{\partial n}G_n(r,\theta;m^{'2})\right|_{n=1}-G_1(r,\theta;m^{'2})\nonumber\\
    &=\sum_{l=0}^{\infty}\sum_{i+j=0}^l \left\{B_{i,j,l}[f](2\partial_{\omega})^i(2\partial_{\bar{\omega}})^j \left[\left.\frac{\partial}{\partial n}\Tilde{G}_n^{\left(1+\frac{i+j+l}{2}\right)}(\mathbf{r},\mathbf{r'})\right|_{n=1}-\Tilde{G}_1^{\left(1+\frac{i+j+l}{2}\right)}(\mathbf{r},\mathbf{r'})\right]\right\}_{\mathbf{r}=\mathbf{r'}}.
\end{align}
For those terms with $i=j$ in the summation, it is easy to figure out by using $Q(x)$, i.e.
\begin{align}
    (4\partial_{\omega}\partial_{\bar{\omega}})^i \left[\left.\frac{\partial}{\partial n}\Tilde{G}_n^{\left(1+i+l/2\right)}(\mathbf{r},\mathbf{r'})\right|_{n=1}-\Tilde{G}_1^{\left(1+i+l/2\right)}(\mathbf{r},\mathbf{r'})\right]_{\mathbf{r}=\mathbf{r'}}=\frac{1}{2\pi}\int_0^{\infty}\frac{-(-\lambda)^{2i+1}}{\left(\lambda^2+m^{'2}\right)^{1+i+l/2}}Q(\lambda r)d\lambda.
\end{align}
However, the remaining terms where $i\neq j$ are still a problem that we can not solve easily. So how to deal with other terms can be a topic for further study.

\appendix
\section{Evaluation of $\int^{\infty}_xJ_0(t)Y_0(t)/t^2 dt$}
The definition of Meijer G function is following:
\begin{align}
   G_{p,q}^{\,m,n}\!\left(z\;\middle\vert\,{\begin{matrix}a_{1},\dots ,a_{p}\\b_{1},\dots ,b_{q}\end{matrix}}\right)={\frac {1}{2\pi i}}\int _{L}{\frac {\prod _{j=1}^{m}\Gamma (b_{j}-s)\prod _{j=1}^{n}\Gamma (1-a_{j}+s)}{\prod _{j=m+1}^{q}\Gamma (1-b_{j}+s)\prod _{j=n+1}^{p}\Gamma (a_{j}-s)}}\,z^{s}\,ds,
\end{align}
where the integral contour $L$ can be found in (9.302) in \cite{GR}. This is a Mellin-Barnes type integral which can be viewed as an inverse Mellin transform. The Mellin transform of this function can be written as
\begin{align}\label{mellin}
    \mathcal{M}\left(G_{p,q}^{\,m,n}\!\left(z\;\middle\vert\,{\begin{matrix}a_{1},\dots ,a_{p}\\b_{1},\dots ,b_{q}\end{matrix}}\right)\right)(s)=\frac {\prod _{j=1}^{m}\Gamma (b_{j}+s)\prod _{j=1}^{n}\Gamma (1-a_{j}-s)}{\prod _{j=m+1}^{q}\Gamma (1-b_{j}-s)\prod _{j=n+1}^{p}\Gamma (a_{j}+s)}
\end{align}
First, we apply Mellin transform and use the multiplicative convolution formula of Mellin transform \cite{tables}
\begin{align}
    \mathcal{M}\left(\int^{\infty}_x\frac{J_0(t)Y_0(t)}{t^2}dt\right)(s)=\mathcal{M}\left(\int^{\infty}_0u\left(1-\frac{x}{t}\right)\frac{J_0(t)Y_0(t)}{t}\frac{dt}{t}\right)(s)\nonumber\\
    =\mathcal{M}\left(u\left(1-x\right)\right)(s)\mathcal{M}\left(\frac{J_0(x)Y_0(x)}{x}\right)(s),
\end{align}
where $u(x)$ is Heaviside step function. From \cite{tables}, we have
\begin{align}
    \mathcal{M}\left(u\left(1-x\right)\right)(s)&=\frac{1}{s}=\frac{\Gamma(s/2)}{2\Gamma(s/2+1)},\nonumber\\
    \mathcal{M}\left(J_0(x)Y_0(x)\right)(s)&=-2^{s-1}\pi^{-1}\frac{\Gamma^2(s/2)\Gamma(1-s)}{\Gamma^2(1-s/2)}\cos(s/2)\pi\nonumber\\
    &=-\frac{1}{2\sqrt{\pi}}\frac{\Gamma^2(s/2)}{\Gamma(1-s/2)\Gamma((1+s)/2)}.
\end{align}
If we set
\begin{align}
    \mathcal{M}\left(J_0(x)Y_0(x)\right)(s)=F(s),
\end{align}
then according to the property of Mellin transform, we have
\begin{align}
    \mathcal{M}\left(\frac{J_0(x)Y_0(x)}{x}\right)(s)=F(s-1).
\end{align}
Thus, the Mellin transform of the original integral turns
\begin{align}
    \mathcal{M}\left(\int^{\infty}_x\frac{J_0(t)Y_0(t)}{t^2}dt\right)(s)=-\frac{1}{4\sqrt{\pi}}\frac{\Gamma^2(s/2-1/2)\Gamma(s/2)}{\Gamma(3/2-s/2)\Gamma(s/2)\Gamma(s/2+1)}.
\end{align}
By using the property of Mellin transform
\begin{align}
    \mathcal{M}\left(f(t^a)\right)(s)=\frac{1}{a}F(\frac{s}{a}),
\end{align}
and comparing to the Mellin transform of the Meijer G function in \eqref{mellin}, it is easy to check
\begin{align}
    \int^{\infty}_x\frac{J_0(t)Y_0(t)}{t^2}dt=G^{20}_{13}\left(x^2\middle\vert
    \begin{array}{c}
         1 \\
         -\frac{1}{2},-\frac{1}{2},-\frac{1}{2}
    \end{array}
    \right).
\end{align}

\section{Derivation of \eqref{full}}
We can derive this form in complex plane. By setting $\omega=x+iy$, the Laplace operator becomes $\nabla^2_{\mathbf{r}}=4\partial_{\omega}\partial_{\bar{\omega}}$. Now the potential $f=f(\omega,\bar{\omega})$ and it is real, i.e. $f=\bar{f}$. The Green's function $G_n(\mathbf{r},\mathbf{r'})$ is the integral kernel of the resolvent $\left(-4\partial_{\omega}\partial_{\bar{\omega}}+m^2-f\right)^{-1}$. 

The expansion of the resolvent operator in powers of $\left(-4\partial_{\omega}\partial_{\bar{\omega}}+m^2\right)^{-1}$ gives
\begin{align}
    \left(-4\partial_{\omega}\partial_{\bar{\omega}}+m^2-f\right)^{-1}=\left(-4\partial_{\omega}\partial_{\bar{\omega}}+m^2\right)^{-1}+\left(-4\partial_{\omega}\partial_{\bar{\omega}}+m^2\right)^{-1}f\left(-4\partial_{\omega}\partial_{\bar{\omega}}+m^2\right)^{-1}+...
\end{align}
Then following the chapter 5 in the paper \cite{dikii1958trace}, the commutator of the resolvent $\left(-4\partial_{\omega}\partial_{\bar{\omega}}+m^2\right)^{-1}$ and $f$ shows
\begin{align}
    &(-4\partial_{\omega}\partial_{\bar{\omega}}+m^2)^{-1}f\nonumber\\
    &\ =f(-4\partial_{\omega}\partial_{\bar{\omega}}+m^2)^{-1}+4(-4\partial_{\omega}\partial_{\bar{\omega}}+m^2)^{-1}(\partial_{\omega}f\partial_{\bar{\omega}}+\partial_{\bar{\omega}}f\partial_{\omega})(-4\partial_{\omega}\partial_{\bar{\omega}}+m^2)^{-1}.
\end{align}
By recurring the commutator, we can finally factor out all $f$ and its derivatives and leave the resolvent $\left(-4\partial_{\omega}\partial_{\bar{\omega}}+m^2\right)^{-1}$ and differential operators $\partial_{\omega}$ and $\partial_{\bar{\omega}}$. Hence, the full resolvent can be expressed as
\begin{align}\label{coef}
    \left(-4\partial_{\omega}\partial_{\bar{\omega}}+m^2-f\right)^{-1}=\sum_{l=0}^{\infty}\sum_{i+j=0}^l B_{i,j,l}[f](2\partial_{\omega})^i(2\partial_{\bar{\omega}})^j \left(-4\partial_{\omega}\partial_{\bar{\omega}}+m^2\right)^{-1-\frac{i+j+l}{2}},
\end{align}
where $i+j+l$ is always even. By acting $\left(-4\partial_{\omega}\partial_{\bar{\omega}}+m^2-f\right)$ on each side, we have the recurrence relations to determine the coefficients:
\begin{align}
    \left\{
    \begin{array}{ll}
         &B_{0,0,0}\equiv1,\ \ B_{i,j,l}=0\ \ \text{for}\ i,j,l<0,  \\
         &B_{i,j,l}=4\partial_{\omega}\partial_{\bar{\omega}}B_{i,j,l-2}+2\partial_{\omega}B_{i,j-1,l-1}+2\partial_{\bar{\omega}}B_{i-1,j,l-1}+fB_{i,j,l-2}.
    \end{array}
    \right.
\end{align}
Also, we need an additional constraint, $B_{i,j,l}=\bar{B}_{j,i,l}$, to keep the resolvent real.

To connect the resolvents and the Green's functions, we can consider function basis in the $n$-sheeted geometry $\{e^{i k/n \theta}J_{|k/n|}(\lambda r)|k\in\mathbb{Z}, \lambda>0,\lambda\in\mathbb{R}\}$. Hence, the Delta function can be constructed
\begin{align}
    \frac{1}{r}\delta(r-r')\delta(\theta-\theta')=\frac{1}{2n\pi}\sum_{k=-\infty}^{\infty}\int_0^{\infty} e^{i k/n (\theta-\theta')}J_{|k/n|}(\lambda r)J_{|k/n|}(\lambda r')\lambda d\lambda.
\end{align}
Then we can plug this expression into \eqref{coef} and we have the form in \eqref{full}
\begin{align}\label{recur}
    G_n(\mathbf{r},\mathbf{r'})=\sum_{l=0}^{\infty}\sum_{i+j=0}^l B_{i,j,l}[f](2\partial_{\omega})^i(2\partial_{\bar{\omega}})^j \Tilde{G}_n^{\left(1+\frac{i+j+l}{2}\right)}(\mathbf{r},\mathbf{r'}),
\end{align}
where
\begin{align}
    \Tilde{G}_n^{\left(1+\frac{i+j+l}{2}\right)}(\mathbf{r},\mathbf{r'})&=\left(\Tilde{G}_n\circ...\circ\Tilde{G}_n\right)(\mathbf{r},\mathbf{r'})\nonumber\\
    &=\frac{1}{2n\pi}\sum_{k=-\infty}^{\infty}e^{i k/n (\theta-\theta')}\int_0^{\infty} \frac{J_{|k/n|}(\lambda r)J_{|k/n|}(\lambda r')}{(\lambda^2+m^2)^{\left(1+\frac{i+j+l}{2}\right)}}\lambda d\lambda.
\end{align}

\bibliography{mybib}
\bibliographystyle{unsrt}

\end{document}